\documentclass[11pt]{article}

\usepackage{bm}

\begin{document}

\title{Space-Time in Quantum Theory}
\date{}

\author{H. Capellmann
\\
Institut f\"{u}r Theoretische Physik, RWTH-Aachen, Germany}

\date{}
\maketitle

\begin{abstract}
{Quantum Theory, similar to Relativity Theory, requires a new concept of space-time, imposed by a universal constant. While velocity of light $c$ not being infinite calls for a redefinition of space-time on large and cosmological scales, quantization of action in terms of a finite, i.e. non vanishing, universal constant $h$ requires a redefinition of space-time on very small scales. Most importantly, the classical notion of ``time'', as one common continuous time variable and nature evolving continuously ``in time'', has to be replaced by an infinite manifold of transition rates for discontinuous quantum transitions. The fundamental laws of quantum physics, commutation relations and quantum equations of motion, resulted from Max Born's recognition of the basic principle of quantum physics:  {\bf To each change in nature corresponds an integer number of quanta of action}.  Action variables may only change by integer values of $h$, requiring all other physical quantities to change by discrete steps, ``quantum jumps''. The mathematical implementation of this principle led to commutation relations and quantum equations of motion. The notion of ``point'' in space-time looses its physical significance; quantum uncertainties of time, position, just as any other physical quantity, are necessary consequences of quantization of action.} 
\\

\end{abstract}


\begin{section}
{Introduction}
\end{section}

When Max Born and Pascual Jordan published the fundamental equations of Quantum Theory in September 1925  \cite{1}, their highly peculiar mathematical form was met with widespread skepticism and misunderstanding. Physical quantities were no longer represented by continuous variables, but by Hermitian matrices; the familiar differential equation of classical physics were replaced by mysterious equations relating different matrices to each other. Before the scientific community had the time to analyze the rationale leading to these equations and their physical content thoroughly, Schr\"{o}dinger published the stationary Schr\"{o}dinger equation \cite{2} in late January 1926, five months later the time-dependent Schr\"{o}dinger equation \cite{3}.  Their mathematical form was more familiar, partial differential equations.  But the central quantity to be determined, the wave-function $\psi ({\bm r},t)$, was equally mysterious.
\\

Both Matrix and Wave Mechanics originated from the conviction that the classical concepts of Bohr's Old Quantum Theory had to be abandoned to understand quantum phenomena; radically new concepts were required.  But Born-Jordan on one side and Schr\"{o}dinger on the other had very different ideas concerning the physical content to be described. They agreed that radical changes away from the foundations of classical physics were required; but the directions into which they proceeded were opposite. Matrix Mechanics was built on the particle concept. That was still similar to Newtonian mechanics. But here the similarity ends; already several years before the final version of Matrix Mechanics was published, Born was convinced that the very foundation of Newtonian physics is no longer valid on atomic and subatomic scales: The space-time continuum as prerequisite for the formulation of all laws of physics is no longer applicable to the quantum scale. All changes in nature proceed by discontinuous and quantized steps, "quantum leaps" ("Quantenspr\"{u}nge").  The  philosophical principle  'Natura non facit saltus' (nature does not make jumps) is no longer valid; a radically new concept of space and time on the quantum scale is required.
\\

Schr\"{o}dinger rejected Born's new physical concept; the classical concept of space-time as given ``a priori'' should be preserved. Continuity in space and time should remain to be the only acceptable way to conceive of processes in nature and should be reflected in the formulation of its fundamental laws. Schr\"{o}dinger's radical change away from Newtonian physics consisted in elimination of the particle concept; on the elementary scale, electrons and all other material objects should be built up from waves.  
\\ 

The physical interpretation of the new quantum laws, which finally did gain widespread acceptance, was not that of the authors  themselves, but the Copenhagen Interpretation of Werner Heisenberg and Niels Bohr.  Heisenberg's  ``reinterpretation (Umdeutung) paper'' \cite{4} was argued to provide the decisive step. The description of the main ideas, problems, and errors of the Copenhagen interpretation is postponed to {\bf chapter (6)}. But before, {\bf chapters (2 - 5)} contain the primary assertion of this paper: Quantization of action in terms of a finite, i.e. non vanishing, universal quantity $h$ indeed requires a redefinition of space-time on atomic and subatomic scales. Only Born's discontinuous quantum physics provides a logically consistent understanding.\footnote{The recent book by the present author \cite{5}, contains a general overview of the development of elementary Quantum Theory, emphasizing Born's decisive contributions.}  
\\
   
\begin{section}
{The Fundamental Principle of Quantum Physics}
\end{section}

Already well before 1925, Born became convinced that the entire system of basic concepts in physics has to be rebuilt radically. The classical space-time continuum as basis for all understanding must be abandoned on the elementary quantum scale.  In December 1919 he wrote to Wolfgang Pauli \cite{6}:
\\
{\it ``For quite some time already I am pursuing this idea, although without success so far: The solution of all quantum problems must be based on very fundamental principles.
One should not transfer the concept of space-time as a four-dimensional continuum from the macroscopic world of common experience to the atomistic world; manifestly the latter requires a different type of manifold".}\footnote{{\it "Gerade diesen Gedanken verfolge ich seit l\"{a}ngerer Zeit, allerdings bisher ohne positiven Erfolg, n\"{a}mlich, dass der  Ausweg aus allen Quantenschwierigkeiten von ganz prinzipiellen  Punkten  aus gesucht werden muss: man darf die Begriffe des Raumes und der Zeit als ein 4-dimensionales Kontinuum nicht  von der makroskopischen Erfahrungswelt auf die atomistische Welt \"{u}bertragen, diese verlangt offenbar eine andere Art von Mannigfaltigkeit als ad\"{a}quates Bild''.}}
\\

In lectures during the winter semester of 1923/24 (published in November 1924 \cite{7}) Born specified what he had in mind:
\\
{\bf 1)} {\it  ``The systematic transformation of classical mechanics into a discontinuous atomic mechanics''.}\footnote{{\it  ``Die systematische Verwandlung der klassischen Mechanik in eine diskontinuierliche Atommechanik''.} The book ``Vorlesungen \"{u}ber Atommechanik, 1. Band'' (Lectures on atomic mechanics, 1st volume) \cite{7}  is primarily devoted to describe Bohr's ``Old Quantum Theory'' and to demonstrate its deficiencies. At the very end on page 341, Born defines the path towards the {\it ``final atomic mechanics'' (``endg\"{u}ltige Atommechanik'')}. Born had used this term in the preface, the intended  ``2nd volume'' should contain the {\it ``endg\"{u}ltige Atommechanik''}. }
\\
{\bf 2)} {\it ``The new mechanics replaces the continuous manifold of} (classical) {\it states by a discrete manifold, which is described by "quantum numbers''}.\footnote{{\it ``Diese neue Mechanik ist dadurch gekennzeichnet, dass an Stelle der kontinuierlichen Mannigfaltigkeit von Zust\"{a}nden eine diskrete Mannigfaltigkeit tritt, die durch ``Quantenzahlen'' beschrieben wird''.} (page 18)}
\\
{\bf 3)} {\it ``Transitions between different states are determined by probabilities''.}\footnote{{\it Wir schreiben jedem \"{U}bergang zwischen zwei station\"{a}ren Zust\"{a}nden eine a priori Wahrscheinlichkeit zu.'}' (page 10)}
\\

Already at that time, Born must have had a precise idea, a fundamental principle, how to implement the program he had defined. This could not be Bohr's Old Quantum Theory, which was constructed from heuristic arguments based on classical concepts. The lectures of 1923/24 \cite{7} demonstrated their narrow limits. As Born had stated in 1919 \cite{6}, something radically new was required.  It was clear to mostly everyone, that quantization of action contained the key; the main question was: Is there a general principle to be drawn, applicable to all physical phenomena?  Born recognized this fundamental principle of quantum physics: As the term ``action'' suggests, the {\bf dynamical} behavior of all physical systems is quantized: At the elementary level, all changes in nature consists of discontinuous steps, ``quantum jumps'' ({\it ``Quantenspr\"{u}nge''}). {\bf All elementary changes correspond to integer numbers of quanta of action. Action variables may only change by integer multiples of $h$.} This general quantization condition provides the  basis for a logically consistent Quantum Theory; all further conclusions are direct consequences of this quantization condition. 
\\

The key is contained in {\it ``discrete manifold described by quantum numbers''}: Different quantum states $n$ and $m$ are to be distinguished by different sets of {\bf integers} $n = (n_1, n_2, n_3,...)$ and $m = (m_1, m_2, m_3,....)$. The integers $n$ characterize the action variables $J_n$ of the corresponding state. Transitions between quantum states $n$ and $m$ correspond to {\bf changes of action variables} $\Delta J_{n,m}$ by  integer multiples of Planck's quantum of action: 

\begin{equation}
\Delta J_{n,m} = ((n_1 - m_1)h, (n_2 - m_2)h, (n_3 - m_3)h,.....).
\end{equation}

Discontinuous behavior of all elementary processes requires a new concept of space-time at the atomic and subatomic level. The most important consequence concerns ``time". Classically, it is assumed that there exists one common time variable $t$ and all changes in nature are continuous in this time variable $t$. The differential equations of motion of classical physics rely on this assumption. Continuity in time suggested that nature behaves deterministically, at least in principle.  For given initial conditions at some point in time, the solution of the differential equations seemingly determine the behavior at any time in the past or future. Discontinuous quantum behavior eliminates the justification for the notion of a continuous time; the classical time variable $t$ has no physical relevance at the quantum scale. Furthermore, the replacement of the differential equations of motion of classical physics by quantum mechanical difference equations eliminates the justification for determinism. Born concludes {\it ``Transitions between different states are determined by probabilities''.} The continuous time of classical physics has to be replaced by an infinite manifold of transition rates for discontinuous and statistical quantum transitions.
\\

\begin{section}
{The Early Born-Einstein Debate}
\end{section}

Born's intention to replace the classical space-time continuum by a discrete manifold did not come from sudden inspiration, but grew out of discussions with Einstein.  Even before Born, Einstein questioned the relevance of the space-time continuum on atomic and subatomic scales.  Not only his contributions to Quantum Theory  of radiation \cite{8,9}, but also to Relativity Theory played a decisive role.
 In the lecture ``On the Theory of Relativity'' \cite{10}, Einstein explains his motives:
\\
 {\it ``The abandonment of certain notions connected with space, time, and motion hitherto treated as fundamentals must not be regarded as arbitrary, but only as conditioned by observed facts..... It is in general one of the essential features of the theory of relativity that it is at pains to work out the relations between general concepts and empirical facts more precisely. The fundamental principle here is that the justification for a physical concept lies exclusively in its clear and unambiguous relation to facts that can be experienced".}\footnote{\it ``Das Aufgeben gewisser bisher als fundamental behandelter Begriffe \"{u}ber Raum, Zeit und Bewegung darf nicht als freiwillig aufgefasst werden, sondern nur als bedingt durch beobachtete Tatsachen...... Es ist \"{u}berhaupt einer der wesentlichsten Z\"{u}ge der Relativit\"{a}tstheorie, dass sie bem\"{u}ht ist, die Beziehungen der allgemeinen Begriffe zu den erlebbaren Tatsachen sch\"{a}rfer herauszuarbeiten. Dabei gilt stets als Grundsatz, dass die Berechtigung eines physikalischen Begriffes ausschlie\ss lich in seiner klaren und eindeutigen Beziehung zu den erlebbaren Tatsachen beruht."}
\\
This same reasoning also raised the question whether the classical concept about space and time could be maintained at atomic scales.  How could the concept of a ``point'' and extremely small distances in space-time be clearly and unambiguously defined by measurements? On scales of common use, rigid rods and clocks could be used to measure lengths and times; on very large and cosmological scales, the light path replaced rigid rods. On subatomic scales, however, these measuring tools failed.
\\
 
Discontinuous behavior was not alien to Einstein, either. In 1905, he had postulated that radiation consists of elementary objects, photons,  which can only be created and absorbed as finite entities \cite{8}.  In late 1916 and early 1917, Einstein's ``Quantentheorie der Strahlung" (Quantum Theory of Radiation) \cite{9} used the photon concept to describe the necessary conditions for thermal equilibrium between matter and radiation. The transfer of energy and momentum between matter and radiation occurs in discontinuous and statistical steps by emission and absorption of photons. The question whether a continuum theory could still be maintained on the quantum scale arose and had to be answered. 
\\

So it is not really surprising that Einstein questioned the relevance of classical space-time concepts on atomic scales already before Born did. Einstein's letter of 1917 \cite{11} testifies that he struggles with the problem of how to maintain a continuum theory and how to define lengths and times on the quantum scale.  
\\
{\it ``Strictly speaking, even the concept of the $ds^2$ evaporates into an empty abstraction, in that $ds^2$ cannot be construed strictly as a measurement result..... If the molecular interpretation of matter is the correct (practicable) one, that is, if a portion of the world must be represented as a finite number of moving points, then the continuum in modern theory contains much too multi-farious possibilities. I also believe that this multifariousness is to blame for the foundering of our tools of description on quantum theory. The question seems to me to be how one can formulate statements about a discontinuum without resorting to a continuum (space-time); the latter would have to be banished from the theory as an extra construction that is not justified by the essence of the problem and that corresponds to nothing ``real"."} \footnote{{\it ``Streng genommen verfl\"{u}chtigt sich auch der Begriff des $ds^2$ in eine leere Abstraktion, indem $ds^2$ nicht strenge als Messresultat aufgefasst werden kann.... Wenn die molekulare Auffassung der Materie die richtige (zweckm\"{a}\ss ige) ist, d. h. wenn ein Teil der Welt durch eine endliche Zahl bewegter Punkte darzustellen ist, so enth\"{a}lt das Kontinuum der heutigen Theorie zu viel Mannigfaltigkeit der M\"{o}glichkeiten. Auch ich glaube, dass dieses zu viel daran schuld ist, dass unsere heutigen Mittel der Beschreibung an der Quantentheorie scheitern. Die Frage scheint mir, wie man \"{u}ber ein Diskontinuum Aussagen formulieren kann, ohne ein Kontinuum (Raum-Zeit) zu Hilfe zu nehmen; letzteres w\"{a}re als eine im Wesen des Problems nicht gerechtfertigte zus\"{a}tzliche Konstruktion, der nichts ``Reales" entspricht, aus der Theorie zu verbannen."}}
\\

Einstein admits that he has no solution; he decides to tentatively retain the continuum as mathematical tool and let eventual success decide its usefulness. 
\\
{\it ``A logically more satisfactory description is obtainable (a posteriori) by relating the theory's more complex individual solutions to observed facts. A standard could then be correlated with a certain type of atomic system that could not claim a privileged position in the theory. Thus a four-dimensional continuum can still be maintained and, in upholding the postulate of general covariance, it then has the advantage of circumventing the arbitrariness in the choice of coordinates".\footnote{{\it ``Eine logisch befriedigendere Darstellung l\"{a}sst sich dadurch (a posteriori) erzielen, dass man die einzelnen komplexeren L\"{o}sungen der Theorie mit Beobachtungsthatsachen in Beziehung setzt. Ein Ma\ss stab w\"{u}rde dann einem Atomsystem von gewisser Art entsprechen, welches in der Theorie keine Sonderstellung beanspruchen k\"{o}nnte. Dabei kann man immer noch an dem vierdimensionalen Kontinuum festhalten und hat dann bei Festalten an dem Postulat der allgemeinen Kovarianz den Vorteil, der Willk\"{u}r einer Koordinatenwahl zu umgehen."}}}
\\

In a lecture ``Geometry and Experience''\ (``Geometrie und Erfahrung'') \cite{12} in January 1921, Einstein again discusses the problem of space-time on the quantum scale. A mathematical, i.e. purely axiomatic, geometry of space-time must be distinguished from ``practical geometry''. Whereas the first constitutes an abstract mathematical formalism, the latter is meant to be a physical science, which includes the possibility of measurements. Whereas mathematics as such is an exact science, its relation to ``physical reality'' should be viewed critically.
While Relativity Theory constitutes the primary topic of the lecture, the problem of space-time on atomic scales is addressed as well. In particular the notions of ``point'' and ``line'' loose their physical significance on subatomic scales.
\\
{\it ``As far as the propositions of mathematics refer to reality, they are not certain; and as far as they are certain, they do not refer to reality..... In axiomatic geometry the words ``point", ``straight line", etc., stand only for empty conceptual schemata. That which gives them content is not relevant to mathematics......All length-measurements in physics constitute practical geometry........It is true that this proposed physical interpretation of geometry breaks down when applied immediately to spaces of submolecular order of magnitude''.\footnote{\it ``Insofern sich die S\"{a}tze der Mathematik auf die Wirklichkeit beziehen, sind sie nicht sicher, und insofern sie sicher sind, beziehen sie sich nicht auf die Wirklichkeit...... Unter ``Punkt", ``Gerade" usw. sind in der axiomatischen Geometrie inhaltsleere Begriffsschemata zu verstehen. Was ihnen Inhalt gibt, geh\"{o}rt nicht zur Mathematik.....Alle L\"{a}ngenmessung in der Physik ist praktische Geometrie.......Die hier vertretene physikalische Interpretation der Geometrie versagt zwar bei ihrer unmittelbaren Anwendung auf R\"{a}ume von submolekularer Gr\"{o}\ss enordnung.''}}
\\

Nevertheless, Einstein does not rule out  that a mathematical field theory might still be of use in quantum physics, even if the mathematical variables do not have their classical significance.
{\it ``The attempt may still be made to ascribe physical meaning to those field concepts which have been physically defined for the purpose of describing the geometrical behavior of bodies which are large as compared with the molecule. Success alone can decide as to the justification of such an attempt, which postulates physical reality for the fundamental principles of Riemann's geometry outside of the domain of their physical definitions. It might possibly turn out that this extrapolation has no better warrant than the extrapolation of the concept of temperature to parts of a body of molecular order of magnitude''.\footnote{\it ``Man kann versuchen, den Feldbegriffen, die man zur Beschreibung des geometrischen Verhaltens von gegen\"{u}ber dem Molek\"{u}l gro\ss en K\"{o}rpern physikalisch definiert hat, auch dann physikalische Bedeutung zuzuschreiben, wenn es sich um die Beschreibung der elektrischen Elementarteilchen handelt, die die Materie konstituieren. Nur der Erfolg kann \"{u}ber die Berechtigung eines solchen Versuches entscheiden, der den Grundbegriffen der Riemannschen Geometrie \"{u}ber ihren physikalischen Definitionsbereich hinaus physikalische Realit\"{a}t zuspricht. M\"{o}glicherweise k\"{o}nnte es sich zeigen, dass  diese Extrapolation ebensowenig angezeigt ist wie diejenige des Temperaturbegriffes auf Teile eines K\"{o}rpers von molekularer Gr\"{o}\ss enordnung.}}
\\

During these years, Born and Einstein met and discussed regularly. And Born explicitly refers to Einstein when he questions  the relevance of precise coordinates at atomic and subatomic scales.
{\it ``Relativity Theory emerged because Einstein recognized the impossibility in principle to determine absolute simultaneity of two events occurring in different locations''.} And he concludes {\it ``The true  laws of nature are determined only by such quantities, which are observable in principle''.}\footnote{\it ``So ist die Relativit\"{a}tstheorie dadurch entstanden, dass Einstein die prinzipielle Unm\"{o}glichkeit erkannte, absolute Gleichzeitigkeit zweier an verschiedenen 0rten stattfindender Ereignisse festzustellen....Ein Grundsatz von grosser Tragweite und Fruchtbarkeit besagt, dass in die wahren Naturgesetze nur solche Gr\"{o}\ss en eingehen, die prinziplell beobachtbar, feststellbar sind."}\cite{13}..... {\it ``If magnitudes lacking this property occur in our theories, it is a symptom of something defective. In order to determine lengths or times, measuring rods and clocks are required. The latter, however, consist themselves of atoms and therefore break down in the realm of atomic dimensions.....it appears justified to give up altogether the description of atoms by means of such quantities as `coordinates of an electron at a given time'.''}\cite{14}
If ``exact'' is taken to have mathematical significance, neither position nor time nor any other physical quantity may be measured or known ``exactly''. 
\\

Both Einstein and Born had reached the conclusion, that the traditional concept of space-time of macroscopic physics cannot simply be transferred to quantum physics.  Basic notions such as a point in space-time and a precise coordinate system are mathematical constructs, but cannot be defined experimentally; arbitrarily small intervals are unmeasurable in principle. The discontinuities occurring in the interaction of radiation with matter indicate that the notion of ``continuum" {\it  ``would have to be banished from the theory as an extra construction that is not justified by the essence of the problem and that corresponds to nothing ``real"."}\cite{11}
Nevertheless, Born and Einstein chose different mathematical routes to attack the quantum puzzle.  Einstein retained differential equations; he accepted that the functions to be determined and the continuous variables used might not have clear physical significance, hoping that {\it  ``(a posteriori) the theory's more complex individual solutions could be related to observed facts"}\cite{11}.
 Born, on the other hand, aimed for a mathematical representation, which should express the discontinuous and statistical behavior of nature explicitly. The space-time continuum should no longer be part of the theoretical formalism. Quite generally, the classical differential equations should be transformed into quantum mechanical difference equations.
\\ 
 
In his letter of Jan 27, 1920, Einstein reacted to Born's suggestion:\cite{15}
\\
{\it ``I do not believe that one must abandon the continuum in order to solve the problem of quanta.... In principle, of course, the continuum could be abandoned. But how one should describe the relative motion of n points  without the continuum?....I believe as before that an overdetermination ought to be sought with differential equations for which the {\it solutions} no longer have continuum properties. But how??''....}\footnote{\it ``Daran, dass man die Quanten l\"{o}sen m\"{u}sse durch Aufgeben des Kontinuums, glaube ich nicht.....Prinzipiell k\"{o}nnte ja das Kontinuum aufgegeben werden. Wie soll man aber die relative Bewegung von n Punkten beschreiben ohne das Kontinuum?....Ich glaube nach wie vor, man muss eine solche \"{U}berbestimmung durch Differentialgleichungen suchen, dass die {\it L\"{o}sungen} nicht mehr Kontinuumscharakter haben. Aber wie??''....}
\\
Similarly several weeks later: \cite{16}
\\
{\it ``I don't believe that the  theory can dispense with the continuum. But my attempts at giving tangible form to my pet idea of interpreting quantum structure through an overdetermination with differential equations refuse to succeed".}\footnote{\it ``Ich glaube nicht, dass die Theorie das Kontinuum wird entbehren k\"{o}nnen. Es will mir aber nicht gelingen, meiner Lieblingsidee, die Quantenstruktur aus einer \"{U}berbestimmung durch Differentialgleichungen zu verstehen, greifbare Gestalt zu geben."}
\\

During the following decades, Einstein will continue his attempts, based on overdetermination of differential equations. His main aim will be a unified field theory, encompassing General Relativity, electromagnetism, and Quantum Theory; without success, however.
\\

\begin{section}
{Mathematical Implementation of the Fundamental Principle}
\end{section}

The transformation of Born's quantization principle into a mathematical theory is contained in three publications. The first by Born in 1924 \cite {16} presented the basic concept.  It was in this paper that Born coined the term ``Quantenmechanik''. Differential equations of classical mechanics are transformed into difference equations of ``Quantum Mechanics''.
\\

{\bf Classically} all physical quantities are represented by continuous variables, the underlying assumption being that all changes in nature occur continuously in space and time. Action-angle ($J_i,w_i$) variables of the classical Hamilton-Jacobi equations \footnote{Full acquaintance with the Hamilton-Jacobi formalism was not commonplace then, nor is it today. Born himself had obtained his doctorate and habilitation in mathematics. His book ``Vorlesungen über Atommechanik, 1. Band'' \cite{7} provided very detailed discussions of the Hamilton-Jacobi formalism, action-angle variables, and canonical transformations. Similarly, ref. \cite{1} contains full details of matrix calculus.}
provide the starting point. A particular advantage is that the transformation from usual variables  and momenta to action-angle variables does not have to be known to define the action variables. 
Each pair $(q_i,p_i)$ of original coordinate $q_i$ and canonically conjugated momentum $p_i$ is associated with its action variable $J_i$, given by 
\begin{equation}
 J_i = \oint p_i \; dq_i\;.
 \end{equation}
  The integral is to be taken at constant energy, not over the classical motion. The action $J_i$ is the new momentum; the new coordinate $w_i$ is the "angle" around the closed path of integration.  All other physical quantities $g$ become functions of the new momenta and coordinates $g = g(J_i,w_i)$.  Due to periodicity in angle variables $w_i$, physical quantities $g$ may be expanded in a Fourier series $g = \sum_\tau g_\tau (J) e^{2\pi i w \tau}$, where $\tau = (\tau_1,\tau_2,...)$, $J = (J_1,J_2,...)$, and $w \tau = \sum_i w_i \tau_i$. 
  \\
  
{\bf Quantum mechanically}, Born's quantization principle requires that only {\bf quantized action intervals} $\Delta J_i = \tau_i h$ (all $\tau_i$ integer) are allowed; replacing the classical differentials $dJ_i$ by discrete action intervals, the classical differential equations are transformed into quantum mechanical difference equations.
Eq. (1) takes the form $\Delta J_i(\tau) = J_i(n) - J_i(n \pm \tau_i) = \mp \tau_i h$. 
\\

Notice the difference between Born's quantization condition and Bohr-Sommerfeld quantization: The latter assumes that the integral $ J_i = \oint p_i \; dq_i $ takes on quantized values $n h$ {\bf within} ``stationary states''; Born's quantization, however, is about {\bf action intervals} of transitions! Born's Quantum Theory is about {\bf quantum dynamics}, not about ``stationary states''. Below, the section about quantum uncertainties will show, that all physical quantities {\bf within} any quantum state cannot have perfectly sharp values; quantum uncertainties of all physical quantities are necessary consequences of quantization of action.
\\

Two papers by Born and Jordan followed. The June 1925 paper \cite{13} applies the explicit discretization procedure to the interaction of radiation with atoms. This paper contains the first fully quantum theoretical treatment of this crucial problem, combining Born's quantized mechanics with Einstein's quantized optics \cite{8,9}. Exchange of energy between atoms and radiation occurs by photon absorption and emission. Einstein's transition probabilities for spontaneous photon emission and field induced absorption and emission are obtained. The September 1925 paper \cite{1} finally arrives at commutation relations and quantum equations of motion.
 \\
 
 \begin{subsection}
 {From Born's Quantization Principle to Commutation Relations}
 \end{subsection}
 
The relation between Born's quantization principle and commutation relations contains the key to understanding Quantum Theory. The formal steps are the following. The discussion is restricted to a single degree of freedom; again, classical differential equations provide the starting point. Continuous variables $p$ and $q$ represent classical momentum and canonically conjugated coordinate; $J$ and $w$ the corresponding action and angle. The Fourier expansions for $p = \sum_{\tau} p_{\tau} (J) e^{2\pi i w \tau}$ and $q = \sum_\tau q_\tau (J) e^{2\pi i w \tau}$ are inserted into the classical definition $J = \oint p\; dq$. The $\oint$ is the integral over $w$ from 0 to 1, yielding $J = 2\pi i\sum_\tau \tau q_\tau p_{-\tau}$. The derivative with respect to $J$ results in $\frac{1}{2\pi i} = \sum_\tau \tau \frac{d}{dJ} q_\tau p_{-\tau}$. 
\\

This classical differential equation is transformed into a quantum mechanical difference equation. The classical differential $dJ$ is replaced by the quantum mechanically allowed discrete action intervals $\tau h$ and the classical Fourier components ($q_\tau$, $p_\tau$) are replaced by ``matrix elements''.\footnote{In their preceding paper  \cite{13}, Born and Jordan defined {\bf ``quantum vectors'' (``Quantenvektoren'')}, which are equivalent to ``matrix elements''.}

\begin{equation}
{ \frac{h}{2 \pi i} = \sum_\tau \bigg(q(n+\tau,n)p(n,n+\tau) - q(n,n-\tau)p(n-\tau,n)\bigg)}.
\end{equation}
The integers $n$ and $n \pm \tau$ characterize different quantum states; the $q(n, n\pm \tau)$ and $p(n, n\pm \tau)$ represent the changes of coordinate $q$ and canonically conjugated momentum $p$, caused by the discontinuous transition from state $n$ to state $n \pm \tau$.\footnote{It is implied that there exists a ``ground state'' corresponding to $n_0 = 0$. Furthermore, action values $J(n)$ cannot take negative values ($J(n) \geq 0$), which implies that  the $q(n,m)$ and $p(n,m)$ containing negative indexes are defined to vanish.} Slightly changing the notation, the general quantization condition takes the form:

\begin{equation}
{\frac{h}{2\pi i} = \sum_m  (p_{nm}q_{mn} - q_{nm}p_{mn})}.
\end{equation}
\\
This is the diagonal element of the commutation relation.\footnote{Concerning the factor $2\pi i$: There is no profound physical reason; the  factor $2\pi i$ is due to the representation of physical quantities by their Fourier coefficients. A mathematical representation of Quantum Theory without the factor $2\pi i$ is perfectly possible.}
Born and Jordan then show that all non diagonal elements vanish. Generalizing to arbitrary numbers of degrees of freedom, the general commutation relations are obtained.

\begin{equation}
\mathcal P \;\mathcal Q - \mathcal Q\;\mathcal P = \frac {h}{2\pi i} \mathcal I,
\end{equation}
where $\mathcal I$ is the unit matrix with elements $\mathcal I_{n,m} = \delta_{n,m}$. 
\\

 General physical quantities $g$ (e.g. $q$ and $p$) are represented by matrices $\mathcal G$ with matrix elements $\mathcal G_{n,m}$. 
 Let state $n$ be represented by  the set of integer quantum numbers $n = (n_1,n_2,...)$, state $m$ by  $m = (m_1,m_2,...$). Transitions between states $n$ and $m$  correspond to quantized action intervals $\Delta J_{n,m} = ((n_1 - m_1) h,(n_2 - m_2)h,....)$.  Non-diagonal matrix elements $\mathcal G_{n,m}$ are related to discontinuous changes of physical quantity $g$ caused by corresponding transitions. Diagonal matrix elements are interpreted as average values; e.g. $\mathcal G_{n,n} = g_n$ is related to the average value of the physical quantity $g$ in the corresponding state. Recall that the general quantization condition refers to {\bf action intervals} of transitions, {\bf not} to any physical quantity within a quantum state. Quantization is about how things change, not about how things are. 
 \\
 
 In summary: The commutation relation, {\it ``the refined quantization condition, which provides the basis for all further conclusions"}\cite{1}\footnote{\it ``die versch\"{a}rfte Quantenbedingung, auf der alle weiteren Schl\"{u}sse beruhen".}, represent the mathematical implementation of the fundamental principle of quantum physics: {\bf To each change in nature corresponds an integer number of quanta of action, independent of the system of reference.}
\\

\begin{subsection}
 {Quantum Uncertainties}
 \end{subsection}

Quantum uncertainties are integral parts of discontinuous quantum physics. Matrix Mechanics originated from Born's conviction that mathematically ``exact'' values of positions and times cannot constitute physically relevant notions. The same reasoning applies to all other physical quantities. The mathematical implementations of discontinuous quantum transitions contained quantum uncertainties from the beginning. Already in 1924, when Born replaced classical differential equations by quantum mechanical difference equations \cite{17}, quantum uncertainties were part of discretization. While action differences of discontinuous transitions were quantized, all physical quantities within any given quantum state were obtained from averaging procedures over classical angle variables and discrete action intervals. And when discontinuous changes of action variables had found their compact form in commutation relations \cite{1}, general uncertainty relations for canonically conjugated quantities followed as mathematical consequence:
The commutation relations require that -- for any quantum state $n$ -- physical quantities cannot have perfectly sharp values: Let $\mathcal Q_{n,n} = q_n$ and $\mathcal P_{n,n} = p_n$ be the average values of two canonically conjugated quantities in state $n$; the product of their mean square deviations has a lower bound imposed by Planck's quantum of action. 

\begin{equation}
\frac{\hbar^2}{4} \leq ((\mathcal Q^2)_{n,n} - q_n^2 ) \cdot ((\mathcal P^2)_{n,n} - p_n^2)
\end{equation}
\\
This inequality is a necessary consequence of Born's quantization condition; its mathematical implementation, the commutation relation, contains the inequality (6) as straightforward mathematical consequence. If ``exact" is understood to have its mathematical significance, then no physical quantity may take on ``exact" values.  A compromise relating the uncertainty of the quantity considered to the uncertainty of its canonically conjugated partner has to guarantee that the inequality (6) is fulfilled; perfectly precise and infinitely imprecise values are excluded. Perfect accuracy of particle position would require infinite momentum uncertainty, implying $(\mathcal P^2)_{n,n}= \infty$, i.e. infinite energy. Similar conclusions forbid other physical quantities to take on perfectly precise values; any assumption of exact value of a physical quantity will invariably lead to conclusions incompatible with the quantum laws themselves.
\\

Similar to position and momentum, time and energy are affected by quantum uncertainties. The universal time of classical physics has no place in discontinuous quantum physics, where an infinite manifold of time scales may be defined via transition rates between two states. The only notion of a specific time associated with a particular state $a$ of a quantum system, is its average lifetime, which is related to the energy uncertainty of state $a$.  A detailed discussion of time in quantum physics is given in the following chapter.
\\

\begin{section}
{Time in Quantum Physics}
\end{section}

\begin{subsection}
{Quantum Equations of Motion}
\end{subsection}

 The most important difference between classical and quantum physics concerns the notion of ``time''. Quite generally time is connected with change: Physical objects change their state as a function of time. The equations governing these changes are the equations of motion. In accord with the classical assumption of continuity in time and space, the classical equations of motion are differential equations, i.e. relations between infinitesimally small changes of physical variables. 
Relying on commutation relations eq.(5) as general quantization principle, Born and Jordan transform the differential equations of classical physics into quantum mechanical difference equations \cite{1}. They describe how general physical quantities $g$, e.g. $p$ or $q$ or any other physical quantity $g(q,p)$, change by discontinuous and statistical quantum transitions
\begin{equation}
\dot {\mathcal G} = \frac { 2\pi i}{h} (\mathcal H \;\mathcal G - \mathcal G\; \mathcal H). 
\end{equation}
The quantum equations of motion do not contain time explicitly; $\dot {\mathcal G}$ is not obtained by usual differentiation of $\mathcal G$ with respect to a continuous time variable.  Eq.(7) is a difference equation, not a differential equation; the right hand side of eq.(7) defines the matrix $\dot {\mathcal G}$. 
The Hamiltonian $\mathcal H(\mathcal Q,\mathcal P)$ determines which transitions are allowed and how physical quantities are affected by the discontinuous quantum transitions. 
\\

There remains the question of the physical significance of the matrix $\dot {\mathcal G}$, and, more generally, 
what type of limiting procedure should relate the new quantum laws to those of classical physics. According to Born's reasoning, discontinuous quantum physics is fundamentally different from supposedly continuous classical physics. While the classical differential equations suggest fully deterministic behavior ``in principle'', discontinuous quantum physics is inherently probabilistic. 
A remark is in order concerning the so-called classical limit of letting $h$ go to zero: The limit $h = 0$ does not exist! 
Recall that commutation relations result from the implementation of the principle: ``Action variables may only change by integer multiples of $h$''. And there is no way that the infinite set of natural numbers may continuously be transformed into a continuum. Similarly, there is no way that inherently statistical behavior may be transformed into fully deterministic behavior continuously. Classical differential equations cannot constitute ``true'' laws of nature, but may only provide approximate descriptions of averages, which ignore the underlying physical discreteness. 
\\

Quite logically, the physical significance of matrix elements $\dot\mathcal G_{n,m}$ is obtained from the requirement, that classical results must be recovered {\bf on average}. The {\bf probability} for a discontinuous change of physical quantity $g$, caused by a transition between quantum states $n$ and $m$, is shown to be proportional to $|\dot {\mathcal G}_{n,m}|^2$. Thereby each particular quantum transition is associated with its particular time scale, defined by ``transition probability per unit time''. The new concept of time, adapted to discontinuous quantum behavior, is contained in an infinite manifold of transition rates. 
\\

Compare to the classical concept of time: A clock consisting of some macroscopic oscillator (e.g. a specific vibration mode of a quartz crystal; or an electromagnetic mode in a microwave cavity) defines time via the number of oscillations per unit time. 
The discrete counting process used to define classical time necessarily introduces finite (classical) uncertainties. As usual in classical physics, it is implicitly accepted that -- in practice -- these finite uncertainties cannot be avoided, but it is assumed that -- in principle -- they may be reduced to be infinitesimally small.
Although not directly relevant for individual quantum systems, the classical time variable defined by clocks may be taken as external parameter, serving as scale for the transition rates of discontinuous quantum behavior. 
It has to be kept in mind, however, that classical time defined by clocks cannot be defined exactly; quantum uncertainties  pose a lower limit to all measurements. Quantization of action guarantees that all physical quantities (time, energy, position, momentum, etc....) are affected by quantum uncertainties.
\\

\begin{subsection}
{General Remarks: Time as Operator and Time as External Parameter}
\end{subsection}

In \textsection 22 of their book Elementare Quantenmechanik (Elementary Quantum Mechanics) \cite{18}, Born and Jordan distinguish between time as external parameter, defined by clocks, on one side, and the concept of time relevant for the discontinuous evolution of individual quantum system on the other:
\\

{\it ``The description of a physical system by a time-dependent Hamilton function, where time is used as external parameter, cannot constitute an exact representation of its physical properties, but only an approximate calculation procedure, which contains fundamental omissions''.\footnote{\it ``Die Beschreibung eines physikalischen Systems durvch eine zeitabh\"{a}ngige Hamiltonfunktion, in welcher die Zeit als Zahlparameter aufgefasst wird, kann nicht eine exakte Darstellung der physikalischen Verh\"{a}ltnisse geben, sondern ist lediglich als ein approximatives Rechenverfahren  zu betrachten, das grunds\"{a}tzliche Vernachl\"{a}ssigungen in sich schliesst.''}}
\\

For closed systems, the classical Hamilton-Jacobi formalism considers energy and time to be canonically conjugated; the negative energy takes on the role of canonical ``momentum'', time its canonically conjugated ``variable''. Born and Jordan conclude: Quantum theoretically for a closed quantum system, time $t$ is not commuting with energy $W$:
\\
{\it ``Classical theory teaches that energy $H = W$ and time are canonically conjugated in closed systems. In analogous and corresponding implementation in Quantum Mechanics the $W, t$ have to be represented by certain non commuting symbols, which are governed by the rules analogous to the canonical commutation relations of $p,q$''.}\footnote{\it ``Die klassische Theorie lehrt nun, dass bei einem abgeschlossenen Systeme die Energie H = W und die Zeit t kanonisch
konjugiert zueinander sind. Bei sinngem\"{a}sser korrespondenzm\"{a}ssiger
\"{u}bertragung in die Quantenmechanik m\"{u}ssen also die
W, t analog den p, q durch gewisse nichtkommutative Rechensymbole
dargestellt werden und Rechenregeln unterworfen sein,
die den kanonischen Vertauschungsregeln der p, q analog sind".}
\\
Similarly, for two interacting quantum system with conserved total energy, time is not commuting with energy:
\\ {\it ``Time is not commuting with energy of these systems (canonically conjugate to it)".}\footnote{\it ``Dann ist die Zeit mit der Gesamtenergie dieser Systeme nicht vertauschbar (kanonosch konjugiert zu ihr)".}
\\

For open systems, however, coupled to surroundings, such that back coupling effects to the surroundings are weak and can be neglected, an external time variable $t$ may be admissible as parameter.
\\
 {\it ``The energy of the partial system may be considered to be approximately commuting with the total energy. Therefore the use of  $t$ as parameter may be justified".}\footnote{\it ``die Energie des Teilsystems [darf] n\"{a}herungsweise als mit der zur Gesamtenergie kanonisch konjugierten Zeit vertauschbar angenommen werden. Dadurch kann der Gebrauch von t als Zahlparameter gerechtfertigt werden."}
\\

Further details are contained in \textsection 61:
\\
{\it ``Already in \textsection 22 it was pointed out, that the time canonically conjugate to the energy of a closed system cannot be represented by a real parameter, but itself constitutes a quantity, which does not commute with other measurable quantities of this system. As far as the time so defined is concerned, it is generally false to state that any other quantity $A(p,q)$ is measured at a specific point in time".}\footnote{\it  ``Es ist schon in \textsection 22 darauf hingewiesen worden, dass die zur Energie eines abgeschlossenen Systems kanonisch konjugierte Zeit nicht durch einen Zahlparameter dargestellt werden kann, sondern selbst eine mit den \"{u}brigen messbaren Gr\"{o}ssen an 'diesem System im allgemeinen nicht vertauschbare Gr\"{o}sse ist. Deshalb kann man, mit Bezug auf die so definierte Zeit, im allgemeinen auch nicht davon sprechen, dass irgendeine Gr\"{o}sse A (p, q) in einem bestimmten Zeitpunkt gemessen wird."}
\\
Again the distinction to open systems with negligibly weak  back coupling to an external system (a ``clock") is stressed; the variable $t$ defined by the external clock may be used as external parameter.
\\
{\it ``This time $t$ may be considered to be defined by a system (a ``clock"), not coupled or very weakly coupled to the system considered".}\footnote{``Diese Zeit t [kann man] als durch ein mit dem betrachteten System gar nicht oder ganz schwach gekoppeltes anderes System (eine ``Uhr") definiert auffassen."}
\\

\begin{subsection}
{Time as Operator}
\end{subsection}

First several preliminary remarks about the notation used below. The symbols $\hat\mathsf t,\; \hat\mathsf E, \; \hat {\mathsf r},\, \hat{\mathsf p} $ are used for time-, energy-, position-, and momentum-operators. Similarly,  $\hat\mathsf q$, $\hat\mathsf p$, $\hat\mathsf g$ indicate operators. The operators might be differential or integral operators. Latin or Greek letters will indicate mathematical variables (not to be mistaken for physical notions), e.g. $t,\; E,\; {\bm r},\; {\bm p}$, denote continuous mathematical variables, representing time-, energy-, position-, and momentum-operators. 
\\

The first representation of energy and time by non commuting symbols is contained in the paper by Max Born and Norbert Wiener \cite{19}. After achieving his original aim of a discretized Quantum Mechanics, Born quickly realized that field theoretical representations are easier to handle mathematically. Even slightly before Schr\"{o}dinger, Born and Wiener replaced 
the discrete mathematical forms of matrix mechanics by field theoretical methods. Matrices are replaced by integral and/or differential operators, and different quantum states are represented by functions of continuous variables. Let us use the symbolic notation $\mathcal G_{mn} = (m|\mathsf {\hat g}|n)$; a general physical quantity $g$ is represented by the operator $\hat\mathsf g$. Operators representing canonically conjugated quantities obey the commutation relations $\hat \mathsf p \hat \mathsf q - \hat \mathsf q \hat \mathsf p =  \hbar/i$. Similarly, the quantum equations of motion takes the form
$ \dot {\hat\mathsf g} = \frac { 2\pi i}{h} (\hat\mathsf H \;\hat\mathsf g - \hat\mathsf g\; \hat\mathsf H)$. 
For closed systems with conserved total energy, time and energy operators fulfill the commutation relation

\begin{equation} 
{\hat \mathsf E \hat \mathsf t - \hat \mathsf t \hat \mathsf E = - \frac{\hbar}{i}}.
\end{equation}
The corresponding time-energy uncertainty relation takes the form

\begin{equation}
\frac{\hbar^2}{4} \leq (({\hat\mathsf t^2})_n - \tau_n^2 ) \cdot (({\hat\mathsf E^2})_n - \epsilon_n^2 ),
\end{equation}
where $\tau_n = (n|\hat \mathsf t|n)$ and $\epsilon_n = (n|\hat \mathsf E|n)$ are average lifetime and energy of state $n$.  $({\hat\mathsf t^2})_n = (n|{\hat\mathsf t^2}|n)$ and $({\hat\mathsf E^2})_n = (n|{\hat\mathsf E^2}|n)$ are average values of the corresponding  squared operators. This relation is independent of the special choice taken how to represent time and energy operators. It is common practice to use the time-energy uncertainty relation to {\bf define} average lifetimes of very short lived particles, e.g. in high energy experiments, by measuring energy uncertainties. Other examples are very small lifetimes of unstable nuclei in heavy ion scattering. Natural linewidths in atomic and molecular spectroscopy are further examples.
\\

Born and Wiener chose the representation  $\hat\mathsf t = t;\;\hat \mathsf E  = - \frac {\hbar}{i} \frac{d}{dt}$. A few weeks later, the representations $\hat\mathsf{\bm r}   =  {\bm r}$ and $\hat\mathsf {\bm p} = \frac {\hbar}{i}\nabla_{\bm r}$ were introduced in Schr\"{o}dinger's ``stationary'' equation \cite{2}. 
Let us call this the ``time-position representation'', which represents quantum states by normalized functions $\psi(t,{\bm r})$ of the continuous variables $t,{\bm r}$. 
\\

A note of caution is in order: The variables $t$ and ${\bm r}$ were introduced as mathematical representations of operators $\hat\mathsf t$ and $\hat\mathsf {\bm r}$.  Instead of $t$ and ${\bm r}$ , we might choose any other symbols; all physically relevant quantities have to be independent of the particular choice taken. In particular, the $t$-dependence of the function  $\psi(t,{\bm r})$ must not be interpreted to represent the continuous evolution as a function of time of an individual physical system.  $\psi(t,{\bm r})$ is nothing more than a mathematical tool, useful to perform calculations of matrix elements.  Physical relevance is contained in averages and probabilities obtained from matrix elements.\footnote{The standard interpretation of  Schr\"{o}dinger's wave function  $\psi(t,{\bm r})$ is different: The variable $t$  is interpreted as classical time variable, and, contrary to Born's understanding,  $\psi(t,{\bm r})$ is interpreted to describe the continuous evolution with time of an individual physical system.
I restate Born's understanding: If $t$ is interpreted as classical time defined by clocks, then $t$ is external parameter and the remark above applies: {\it ``The description of a physical system.....cannot constitute an exact representation of its physical properties, but only an approximate calculation procedure, which contains fundamental omissions.''} The section ``Time as external parameter'' discusses this point in detail.}
\\

In the chosen representation, the equation of motion $ \dot {\hat\mathsf g} = \frac { 2\pi i}{h} (\hat\mathsf H \;\hat\mathsf g - \hat\mathsf g\; \hat\mathsf H)$ is a differential equation with respect to $t$. Formal integration yields:

\begin{equation}
\hat\mathsf g(t)  = e^{\frac {i}{\hbar}\hat H t} \; \hat\mathsf g \; e^{- \frac {i}{\hbar}\hat H t}.
\end{equation}
But again, the note of caution already mentioned above applies: Do not mistake the mathematical variable $t$ to indicate continuous physical behavior. The variable $t$ has been  introduced as representation of the time operator. Mathematically, the commutation relations may be satisfied by infinitely many different representations; no physically relevant and observable quantity may depend on a specific representation. All of physics is contained in probabilities and averages obtained from matrix elements, which are independent of the particular representation used to obtain them.
\\

\begin{subsection}
{Energy-Momentum Representation}
\end{subsection}

Section 4.1 showed that the requirement of discontinuous action intervals resulted in commutation relations. The following example demonstrates that the reverse conclusion holds as well; commutation relations indeed imply discontinuous quantum behavior. The essential differences between classical and quantum behavior are particularly perceptible, if the interactions between two quantum systems become small, such that lowest order perturbation theory is applicable. Classically, small interactions cause small changes in physical quantities; quantum mechanically even small interactions may cause large changes in physical quantities, the corresponding probabilities tend to vanish for vanishing interactions.
\\

As example, I discuss scattering processes of particles by crystals.\footnote{Scattering processes of particles by atoms played a crucial role in Born's statistical interpretation of the wave function \cite{20} Assuming the atomic spectrum to be given, he used wave functions to compute the relevant matrix elements.} {\bf Classically}, the  interaction between particle and crystal  is taken to be a time dependent potential $V({\bm r}, t)$, where the $t$ dependence describes the classical crystal dynamics. Fourier transformation  $V({\bm r}, t) = \int d \omega \; d^3 {\bm k} \;\tilde V(\omega,{\bm k}) \;e^{ i (\omega t - {\bm k} {\bm r})}$ will be helpful in the following. For the {\bf quantum mechanical} treatment, I consider the combined system of particle and crystal to constitute a closed system. Following the Born-Jordan perspective, all physical quantities --  energy, time, momentum, position -- are represented by operators. The commutation relations for time-energy and position-momentum may equally be satisfied by the ``energy-momentum representation''\footnote{The ``energy representation'' was introduced by the present author in the appendix of \cite{5}. The energy operator is represented by the continuous variable $E$ and the time operator by the derivative with respect to $E$.} 

\begin{equation}
\hat \mathsf {E}   =  E;\;\;\;\hat\mathsf t  =  \frac {\hbar}{i} \frac{d}{dE};\;\;\; \hat \mathsf {\bm p}   =  {\bm p};\;\;\;\hat\mathsf {\bm r} = - \frac {\hbar}{i}\nabla_{\bm p}.
\end{equation}

Replacing the  classical variables ${\bm r}$ and $t$ by the operators $\hat\mathsf {\bm r} = - \frac {\hbar}{i}\nabla_{\bm p}$ and $\hat\mathsf t  =  \frac {\hbar}{i} \frac{d}{dE}$, the classical interaction $V({\bm r}, t) = \int d \omega \; d^3 {\bm k} \;\tilde V(\omega,{\bm k}) \;e^{ i (\omega t - {\bm k} {\bm r})}$ becomes the interaction operator $\hat\mathsf V = \int d \omega d^3 {\bm k}\;\tilde V(\omega,{\bm k}) \;e^{ \hbar (\omega \frac{d}{dE} - {\bm k \nabla}_{\bm p})}$. 
The particle is taken to be structureless and without internal dynamics. In the chosen representation, the particle state may be represented by the function $f(E,{\bm p})$.
 The  interaction operator $\hat\mathsf V $ is applied as perturbation to the uncoupled free particle state $f(E,{\bm p})$. {\bf In lowest order} we obtain

\begin{equation}
\int d \omega d^3 {\bm k}\;\tilde V(\omega,{\bm k}) \;e^{ \hbar (\omega \frac{d}{dE} - {\bm k \nabla_{\bm p}})} \; f (E, {\bm p}) = \int d \omega d^3 {\bm k}\;\tilde V(\omega, {\bm k})  \; f (E + \hbar \omega, {\bm p} + \hbar {\bm k}).
\end{equation}
The physical interpretation is obtained from the quantum equation of motion: 
The Fourier component $\tilde V(\omega, {\bm k})$ may cause 
 a discontinuous energy transfer $\Delta E = \hbar \omega$ and momentum transfer $\Delta {\bm p} = \hbar {\bm k}$. The transition probability is proportional to $|\tilde V(\omega, {\bm k})|^2$ \cite{20}. For $\omega \neq 0$, the sign used in the Fourier transform is chosen such that positive $\omega$ correspond to energy transfer from the crystal to the particle (e.g. by absorption of a phonon); negative $\omega$ to energy transfer from the particle to the crystal (e.g. creating a lattice excitation). 
 \\
 
 Only discrete energy and momentum transfers are allowed, because their corresponding action variables may only change by integer multiples of $h$. The symmetries of the interaction potential provide the relation between action intervals $\Delta J_E$ and $\Delta J_{\bm p}$ and allowed energy and momentum transfers $\Delta E$  and $\Delta {\bm p}$. 
The Fourier component $\tilde V(\omega, {\bm k})$ is invariant under translation in time by $t_\omega = \frac{2\pi}{\omega}$ and translation in space in direction of ${\bm k}$ by the length $\frac{2\pi}{|\bm k|}$. From the definition of the action variables, $ J_i = \oint p_i \; dq_i$, it follows that the product of periodicity time $t_\omega$ and allowed energy transfer $\Delta E$ has to be equal to the action interval $\Delta J_E $. Similarly, the allowed momentum transfers are in direction of symmetry vector ${\bm k}$, the product of their magnitude  and periodicity length $\frac{2\pi}{|\bm k|}$ has to be equal to the corresponding action interval $\Delta J_{\bm p}$.
\begin{equation}
\Delta J_E = \Delta E \; \frac {2\pi}{\omega}  = h;\;\;\;\; \Delta J_{\bm p} = |\Delta {\bm p}| \; \frac{2\pi}{|\bm k|}= h.
\end{equation}
In lowest order perturbation theory, the change in action variables due to any one of the various Fourier component $\tilde V(\omega,{\bm k})$ is equal to the smallest value allowed by the fundamental laws, Planck's quantum of action $h$. Strong interactions may lead to changes of action variables by multiple integers of $h$; two photon absorption in strong electric fields is an example.
\\

Time and energy have been represented by non-commuting symbols, implying that total energy of particle and crystal combined is conserved. Implicitly, the crystal is treated quantum theoretically, too. A discrete transition of the particle changing its energy and momentum has to be coupled to a discrete transition in the crystal. This fact is used experimentally, to study the elementary excitations of crystals, e.g. lattice excitations (phonons)  or magnetic excitations (magnons). 
\\ 

These results were obtained in {\bf lowest order perturbation theory}. Classically, lowest oder is adequate for infinitesimally small perturbations causing equally infinitesimally small changes in the perturbed system. For vanishing strength of the interaction, the classical response tends towards zero continuously. The quantization condition, implemented by the commutation relations, predicts very different results. Small, even extremely weak perturbations may cause large and discontinuous changes of physical quantities; the corresponding {\bf probabilities} tend towards zero for vanishing strength of the perturbation. Furthermore, if the interaction is weak enough, such that lowest order effects only have to be considered, all types of scattering potentials may be treated. Decomposing the scattering potential into its different Fourier components, the total scattering probability is obtained from the sum over the contributions of the various Fourier coefficients. 
\\
 
Elastic transitions ($\Delta E = 0$ and $\Delta J_E = 0$), i.e. the scattering contributions due to Fourier components $\tilde V(\omega = 0, {\bm k})$, are of particular interest for diffraction phenomena. 
Systems of discrete translational symmetries produce particularly large scattering probabilities for special momentum transfers $\Delta {\bm p_i} = \hbar {\bm Q_i}$, the ``Bragg peaks''. In crystalline materials a large fraction of the total elastic scattering intensity is concentrated in Bragg scattering, easily distinguishable from the usually structureless background of inelastic processes.
The special momentum transfers contributing to Bragg scattering are representative of translational symmetries; translation in  direction of any one of the ${\bm Q_i}$ by the corresponding length $\frac{2\pi}{|\bm Q_i|}$ represents a crystalline symmetry operation. Crystallography relies on this correspondence; identification of a large enough number of Bragg peaks permits the identification of crystal structure and interatomic distances.
\\

Remark that the occurrence of Bragg peaks does not rely on any intrinsic wave property of the scattered particles.  The  special momentum transfers $\Delta {\bm p_i}$ in Bragg scattering are due to the fundamental quantization condition $ \Delta J_{\bm p} = |\Delta {\bm p_i}| \; \frac{2\pi}{|\bm Q_i|}= h$, not to any alleged wave property of the scattered particles.
\\

\begin{subsection}
{Time as External Parameter, Time Dependent Perturbation Theory}
\end{subsection}

Let us recall Born's remarks, concerning the use of explicitly time dependent Hamiltonians:
{\it ``The description of a physical system by a time-dependent Hamilton function, where time is used as external parameter, cannot constitute an exact representation of its physical properties, but only an approximate calculation procedure, which contains fundamental omissions''.\footnote{\it ``Die Beschreibung eines physikalischen Systems durch eine zeitabh\"{a}ngige Hamiltonfunktion, in welcher die Zeit als Zahlparameter aufgefasst wird, kann nicht eine exakte Darstellung der physikalischen Verh\"{a}ltnisse geben, sondern ist lediglich als ein approximatives Rechenverfahren zu betrachten, das grunds\"{a}tzliche Vernachl\"{a}ssigungen in sich schliesst.''}}\cite{18}.
\\

``Time-dependent'' perturbation theory is a standard example, where the time variable $t$  must be treated as external parameter.
An explicitly time dependent perturbation $V(t) = \int d \omega \tilde V(\omega) e^{i\omega t}$ is applied to a quantum system. Just as in the preceding section, lowest order effects only are considered. At $t = 0$ the system is taken to be in the state $\phi_i$ of energy $\epsilon_i$; under the influence of the time dependent external perturbation $V(t)$, the time dependent Schr\"{o}dinger equation predicts $\phi_i$ to develop into $\psi (t) = \sum_\nu c_\nu (t) \phi_\nu$. Typical textbooks identify $|c_\nu (t)|^2$ as ``probability to find the system at time $t$ in the state $\phi_\nu$''. Detailed calculations  are textbook material, I directly address the physically relevant result, ``Fermi's Golden Rule'': In the limit of large $t$, the transition probability from state $\phi_i$ of energy $\epsilon_i$ to the state $\phi_f$ of energy $\epsilon_f$ is proportional to $|\tilde V(\omega)|^2 \delta (\epsilon_f - \epsilon_i +\hbar \omega) \; t$. The limit of large $t$ has led to the delta-function, guaranteeing energy conservation. The transition rates  $|\tilde V(\omega)|^2 \delta (\epsilon_f - \epsilon_i +\hbar \omega)$ reproduce the results obtained by the (much simpler) method of the preceding section.\\

Only these transition rates are physically relevant and experimentally measurable. 
Typical textbook claims, that $\psi (t)= \sum_\nu c_\nu (t) \phi_\nu$ describes the temporal evolution of the state of an individual quantum system for arbitrary values of ``time'' $t$,  are incorrect.  It impossible in principle to back up these claims experimentally; time-energy quantum uncertainties pose a lower limit to time and energy accuracies. 
\\

Although the $t$-dependent wave function does not describe the continuous evolution of an individual physical system, $\Psi (t)$ does constitute an {\bf approximate} description, provided that the variable $t$ is interpreted as external parameter, defined by a classical clock, and $\Psi(t)$ is interpreted as representation of an ensemble, i.e. a very large number of equivalent quantum systems. At $t =0$, the ``system'' to be described by $\Psi(t=0)$ consists of a macroscopically large number $N$ of equally prepared quantum systems in the initial state $\phi_i$.  According to the time dependent Schr\"{o}dinger equation, $\Psi (t=0)$ develops into $\Psi (t) = \sum_\nu c_\nu (t) \phi_\nu$. Applying the Born(-Einstein) ensemble interpretation (described extensively in the following section), $| c_\nu (t)|^2 = n_\nu (t)$ for $\nu \neq i$  then is the number of individual quantum system having made a transition from the intial state $\phi_i$ to the state $\phi_\nu$, where $\sum_\nu n_\nu (t) = N$. To lowest order in $\frac{1}{N} \sum_{\nu \neq i} n_\nu (t)$, the result given by Fermi's golden rule reproduces the transition rate of Born's statistical interpretation, described extensively in the following section.
\\

\begin{subsection}
{The Born(-Einstein) Ensemble Interpretation}
\end{subsection}

When Born derived the statistical interpretation of the wave function in 1926 \cite{20}, there was no time variable $t$ involved. He obtained the transition probabilities directly from matrix elements, which he determined using time independent perturbation theory. Nevertheless, ref. \cite{20} indicates how $\psi(t,{\bm r})$ may constitute an {\bf approximate procedure}, provided that the variable $t$ is interpreted as external parameter, and $\psi(t,{\bm r})$ is not interpreted as representation of a single quantum system, but of an ensemble, i.e. a very large number of equivalent quantum systems.
The name of Einstein is included in parenthesis in the title above, because Einstein adopted the ensemble interpretation of $\psi (t)$ \cite{21} in exactly the sense intended by Born.
\\

Born's statistical interpretation \cite{20} consists of two papers, the ``Preliminary Announcement" of June 1926 was followed by the extended version one month later. The subject of the two papers is indicated by their title: ``Quantum Mechanics of Collision Processes" ({\it ``Quantenmechanik der Sto\ss vorg\"{a}nge"}): Particles (e.g. electrons) are scattered by atoms. Consistent with the basic postulate of discontinuous quantum physics, there is no explicit time variable. The aim consists in calculation of transition matrix elements, which, in turn, will determine transition probabilities.
The initial state consist of separate quantum systems, particles and atoms, far apart from each other and  noninteracting. Particles and atoms collide; asymptotically in the final state, particles and atoms again are far apart from each other and noninteracting.\footnote{This experimental situation corresponds precisely to the configuration addressed in the EPR-(Einstein-Podolsky-Rosen) paper of 1935 \cite{22}, which stimulated Schr\"{o}dinger's papers about entanglement and ``Schr\"{o}dinger's Cat"\cite{23}. 
The ``EPR"- paper actually was not written by Einstein, but by  Podolsky and does not constitute Einstein's own views properly. Einstein's own opinion on this matter is contained in Einstein's review ``Physics and Reality" of 1936 \cite{21}. A detailed account is given by Arthur Fine in ``The Shaky Game, Einstein's Realism and the Quantum Theory" \cite{24}. Further details are given in the present author's book \cite{5}. } 
\\

Born's preliminary scattering paper \cite{20} describes the collision of a single electron with an atom. The premise is that{\it ``before, as well as after, the collision, when the electron is far away and the coupling small,  a particular state of the atom and a particular, rectilinear-uniform movement of the electron has to be definable".}\footnote{\it ``sowohl vor als auch nach dem Sto\ss e, wenn das Elektron weit genug entfernt und die Koppelung klein ist, [muss] ein bestimmter Zustand des Atoms und eine bestimmte, geradlinig-gleichf\"{o}rmige Bewegung des Elektrons definierbar sein.''}
Mathematically, the task is reduced to determine the asymptotic behavior produced by the collision.
Born distinguishes between ``physical" (or ``real") states of electrons and atoms and their mathematical representations. Particles are represented by plain wave functions of wave vector ${\bm k}$.  Physical significance (or ``reality") is attributed to electron momenta ${\bm p} = {\hbar{\bm k}}$ and atom energies $\epsilon$ only. 
Wave functions serve as mathematical tools to calculate transition probabilities from real initial states to real final states. 
The initial noninteracting state is represented in terms of product wave functions $\Psi_i = \psi_{{\bm k}_i} \psi_m$ of free electron $\psi_{{\bm k}_i}$ with momentum $\hbar {\bm k}_i$ and free atom  $\psi_m$ with energy $\epsilon_m$. The transition probabilities from initial to various final states caused by the (weak) electron-atom interaction are calculated.  The first short paper does not contain mathematical details, but simply states the essential result and Born's interpretation of its physical significance. Asymptotically, after the collision has taken place, the total $\Psi$-function takes the form
\\

\begin{equation}
\Psi_s= \sum_n \int d^2 \Omega_{{\bm k}_f} \; \Phi ({\bm k}_i m, {\bm k}_f n) \; \psi_{{\bm k}_f} \psi_n. 
\end{equation}

\noindent
The integral over $ d^2 \Omega_{{\bm k}_f}$ is over the solid angle of outgoing momenta ${\hbar\bf k}_f$.  Energy conservation  ($\epsilon_m - \epsilon_n$ =  $\hbar^2 |{\bm k}_f|^2 / 2 m_e$ - $\hbar^2 |{\bm k}_i|^2 / 2 m_e$) determines the absolute value $|{\bm k}_f|$.
\\ 

The wave function $\Psi_s$ consists of a superposition of many product wave functions $\psi_{{\bm k}_f} \psi_n$. Born concludes:
{\it ``We do not get an answer to the question, ``what is the state after the collision?'', but only to the question, ``how probable is a given result of the collision?''... based on the principles of our Quantum Mechanics there exists no quantity, which determines the result of the collision for the individual elementary process''.}\footnote{\it Man bekommt keine
Antwort auf die Frage, ``wie ist der Zustand nach dem Zusammensto\ss e'', sondern nur auf die Frage, ``wie wahrseheinlieh ist ein vorgegebener Effekt des Zusammensto\ss es''.....Vom Standpunkt unserer Quantenmechanik gibt es keine Gr\"{o}\ss e, die im Einzelfalle den Effekt eines Sto\ss es kausal festlegt.''}
The physical significance of the superposition $\Psi_s= \sum_n \int d^2 \Omega_{{\bm k}_f} \; \Phi ({\bm k}_i m, {\bm k}_f n) \; \psi_{{\bm k}_f} \psi_n$ is reduced to: 
\\

The transition probability from initial state represented by $ \psi_{{\bm k}_i} \psi_m$ (electron with momentum ${\hbar{\bm k}_i}$ and atom with  energy  $\epsilon_m$) to any one of the possible final states represented by $\psi_{{\bm k}_f} \psi_n$ (electron momentum ${\hbar{\bm k}_f}$ and atom energy $\epsilon_n$)   is proportional to the absolute square   $|\Phi ({\bm k}_i m, {\bm k}_f n)|^2$.
\\

The same reasoning is adopted for light scattering, which, based on Einstein's concept of photons, should be understood as particle scattering: {\it ``I further believe that the problem of absorption and emission of light must also be
treated in a completely analogous way...... in accord with the concept of light quanta."}
\\

The extended version published one month later replaces the rather academic problem of colliding one electron with one atom by the typical physical situation realized in the laboratory: A stationary current of particles (electrons) is produced and brought into collision with a gas consisting of a large number of atoms. Again wave functions are used as mathematical tools to calculate the transition probabilities. 
Born specifies his general interpretation of Schr\"{o}dinger's $\psi$-functions. He refers to Einstein's expression of  "ghost field" ({\it ``Gespensterfeld"}) \cite{25}, introduced in the context of light scattering. Whereas energy and momentum are carried by particles (i.e. photons), the fictitious ghost field describes the probability distribution over large numbers of single particle scattering events. Schr\"{o}dinger's wave functions are similar ghost fields without direct physical significance. Momentum and energy are transferred in such a way that
{\it ``particles (electrons) actually fly about (als wenn Korpuskeln (Elektronen) tats\"{a}chlich (= actually, really, in fact) herumfliegen''}). 
{\it The flight path of the particles is determined only in so far, as restricted by energy and momentum conservation; apart from that, the probability for a particular path is governed by the function $\psi$. Particle dynamics are determined by probability laws".}\footnote{\it ``Die Bahnen dieser Korpuskeln sind nur so weit bestimmt, als Energie und
Impulssatz sie einschr\"{a}nken; im \"{u}brigen wird f\"{u}r das Einschlagen
einer bestimmten Bahn nur eine Wahrscheinlichkeit durch die Werteverteilung
der Funktion $\psi$ bestimmt. Die Bewegung der Partikeln folgt Wahrscheinlichkeitsgesetzen.''}
\\

The interpretation of the ghost field $\psi$ is adapted to the physical problem, i.e. a large number (or ``ensemble") of atoms and particles. The representation of an ensemble of noninteracting atoms is given in terms of the eigenfunctions $\psi_n(q)$ with eigenvalues $\epsilon_n$  of the stationary Schr\"{o}dinger equation. Since the system of functions $\psi_n(q)$ is complete, any function $f(q)$ may be expanded in terms of the eigenfunctions $f(q) = \sum_n c_n \psi_n(q)$. 
Born asks the question: If the normalized functions $\psi_n(q)$ constitute representations of atomic states of energy $\epsilon_n$, what type of physical system might be associated with superpositions? 
Born's conclusion is the following:
A superposition $f(q) = \sum_n c_n \psi_n(q)$ is related to the  {\it ``probability for the occurrence of the various states in a mixture of equal and uncoupled atoms. The completeness relation $\int dq \; |f(q)|^2 = \sum_n |c_n|^2$ leads to regard this integral as the number of the atoms..... $|c_n|^2$  denotes the abundance of the state $n$ and the total number is composed of the sum over the various contributions".}\footnote{\it ``Wahrscheinlichkeit daf\"{u}r, dass in einem Haufen gleicher, nicht gekoppelter Atome die Zust\"{a}nde in einer bestimmten H\"{a}ufigkeit vorkommen. Die Vollst\"{a}ndigkeitsrelation  $\int dq \; |f(q)|^2 = \sum_n |c_n|^2$ f\"{u}hrt dazu, dieses Integral als die Anzahl der Atome anzusehen.....$|c(n)|^2|$ bedeutet die H\"{a}ufigkeit des Zustandes n, und die gesamte Anzahl setzt sich aus diesen Anteilen additiv zusammen."} 
\\
In short: $f(q) = \sum_n c_n \psi_n(q)$ is not to be associated with one individual atom but with a mixture (``ensemble") of many atoms and $|c_n|^2$ is to be interpreted as the number of atoms in the state $n$.
\footnote{In 1936 Einstein will refer to this example to illustrate his ``ensemble interpretation" \cite{5}.}
\\

The equivalent reasoning is applied to ensembles of free particles. Any general function $g({\bm r})$ may be expanded in terms of free particle eigenfunctions $\psi_{\bm k}$ (i.e. a simple Fourier expansion). But again a superposition will, in general, {\bf not} represent an acceptable physical state of a single individual particle. For the physical problem under consideration, i.e. an ensemble of free particles: The superposition $g({\bm r})= \sum_{\bm k} c_{\bm k} \psi_{\bf k }$ is related to the probability for the occurrence of various free particle states of momentum $\hbar {\bm k}$ in a mixture (ensemble). The absolute square of the expansion coefficients $|c_{\bm k}|^2$ is to be interpreted as providing the abundance of particles in states of momentum $\hbar {\bf  k }$. 
\\ 

The problem to be solved is specified as follows:\\
{\it ``For the processes considered, the paths of the particles before and after the collision are asymptotically rectilinear. For  a very long time (in comparison with the actual collision process) the particles are in practically free states. In agreement with the experimental situation, we are led to the following approach: Let the distribution function $|c (k)|^2$ for the asymptotic paths before the collision be known; are we able to calculate the distribution function after the collision? Of course, we are considering a stationary current of particles".}\footnote{\it ``Bei diesen Vorg\"{a}ngen hat jede Bewegung vor und nach dem Stosse eine geradlinige Asymptote. Die Teilchen befinden sich also sehr lange (im Vergleich zur eigentlichen Stossdauer) vor und nach dem Stosse in praktisch freiem Zustande. Man kommt daher in \"{U}bereinstimmung mit der experimentellen Problemstellung zu folgender Auffassung:  F\"{u}r die asymptotische Bewegung vor dem Stosse sei die  Verteilungsfunktion $|c (k)|^2$ bekannt; kann man daraus die Verteilungsfunktion nach dem Stosse berechnen ? Dabei ist nat\"{u}rlich hier von einem station\"{a}ren Teilchenstrom die Rede."}
\\

In lowest order, the coefficients  $\Phi ({\bm k}_i m, {\bm k}_f n)$, obtained in the preliminary communication are shown to be proportional to the matrix elements $( \psi_{{\bm k}_i} \psi_m | V_{e.a} | \psi_{{\bm k}_f} \psi_n )$, where $V_{e.a}$ is the electron-atom interaction.\footnote{Higher order contributions are contained in what is usually called the ``Born series''.}
The transition probabilities from  initial states  $\Psi_i = \psi_{{\bm k}_i} \psi_m$ to final states $\Psi_f = \psi_{{\bm k}_f} \psi_n$ are proportional to the absolute squares $| ( \psi_{{\bm k}_i} \psi_m | V_{e.a} | \psi_{{\bm k}_f} \psi_n ) |^2$. 
\\

To summarize: Born's statistical interpretation is about {\bf ``transition probabilities"} of discontinuous and statistical transitions; their probabilities are proportional to absolute squares of off-diagonal matrix elements.  If Wave Mechanics, in addition to Matrix Mechanics, introduces functions $\psi ({\bm r},t)$, these wave functions do not contain any additional physics. In particular, they do {\bf not} describe the continuous evolution in space and time of an individual quantum system. Their usefulness is limited to mathematical tools for computation of matrix elements. No additional physical reality  is to be associated with wave functions; they are nothing more than ghost fields, or phantoms of the imagination.\\

 \begin{section}
{The Copenhagen Interpretation}
\end{section}

Although Matrix Mechanics is generally recognized to contain the basic equations of Quantum Theory, Born's underlying physical concept of discontinuous quantum physics is not. Instead, the Copenhagen interpretation became the dominant way to interpret the new quantum laws. This chapter describes its main postulates, problems, and errors. 
\\

Heisenberg and Bohr are the main architects of the Copenhagen interpretation; the unconditional support by Wolfgang Pauli was essential for its widespread acceptance. Bohr was the spiritual leader; although he did not contribute to the mathematical development of the new quantum laws, his supposedly deeper insight determined the guiding line for his young collaborators. 
During the twelve years preceding the arrival of the new quantum laws, Bohr had established himself as the highest authority of Quantum Theory. His ``Old Quantum Theory'' was based on essentially classical concepts. The laws of Newtonian physics were supplemented by a number of  heuristic rules (``principles''), which selected ``stationary states'' of electrons circling the nucleus. Radiation was supposed to retain its classical character. The formal basis for the Copenhagen interpretation of Quantum Theory was provided in two papers by Heisenberg, the ``re-interpretation paper'' \cite{4} of 1925 and the ``indeterminacy paper'' \cite{26} of 1927. Bohr's  review of 1928 \cite{27} (published simultaneously in Naturwissenschaften and Nature) provided the final touches. Heisenberg had been Born's collaborator in G\"{o}ttingen; during the crucial time of the discovery of the Matrix Mechanics in 1924/25, however, Heisenberg spent most of his time collaborating with Bohr in Copenhagen. He retained the essentially classical mode of thought of Bohr's Old Quantum Theory;  similar to Bohr, Heisenberg considered the final product of Matrix Mechanics to constitute a mathematical formalism only.
\\

\begin{subsection}
{The Doctrine of Classical Concepts} 
\end{subsection}

Bohr's imperative assertion, that classical concepts have to be maintained to describe quantum physics, defined the framework. His old Quantum Theory had relied on classical concepts; continuity of all physical processes in Newtonian  space and time were central elements. Atoms were viewed as miniature solar systems, electrons in stationary states of quantized energies $\epsilon_n$ performing continuous orbits around nuclei. Classical equations of motion provided the energies. Bohr's cardinal error was the rejection of Einstein's quanta of radiation; Bohr maintained that radiation had to retain its classical character, completely described by classical Maxwell equations. But if atoms could emit continuous radiation, then there had to be something oscillating inside the atoms, providing the required frequencies. This role was attributed to ``virtual oscillators''; each spectral line of frequency $\nu$ was associated with its corresponding virtual oscillator.  Bohr's frequency condition $h \nu = \epsilon_n - \epsilon_m$ was based on this assumption. When the new quantum laws were published by Born and Jordan \cite{1}, Bohr's efforts concentrated on constructing an interpretation in accord with the classical concepts of his own old Quantum Theory. 
\\

 \begin{subsection}
{Heisenberg's Re-Interpretation}
\end{subsection}

The origin of Matrix Mechanics was attributed to Heisenberg's ``Quantum Theoretical Re-Interpretation of Kinematic and Mechanical Relations'' of July 1925 \cite{4}. Heisenberg's line of thought still followed the classical lines of Bohr's old Quantum Theory, and he intended to provide a new kinematic description of physical quantities for Bohr's physical picture.
\\

In fact, what Heisenberg really re-interpreted was Born's preceding papers \cite{17} and \cite{13}. The June 1925 paper by Born and Jordan \cite{13} had applied Born's concept of discontinuous quantum physics \cite{17} to the interaction of atoms with radiation; emission and absorption of photons were combined with discontinuous changes of atomic properties. Heisenberg, after his return from Copenhagen, witnessed the final stages of its genesis. His own re-interpretation paper of July 1925 \cite{4} picked up the same subject, but instead of following Born's line of thought, he twisted Born's intentions to signify the contrary. 
He applied a slightly modified Bohr-Sommerfeld quantization procedure to Bohr's virtual oscillators, which supposedly  were responsible for the emission of continuous radiation. The virtual oscillator amplitudes -- in Heisenberg's opinion -- determined the intensities of the corresponding spectral lines.
\\

 The essential differences and similarities between the Born-Jordan paper of June 1925 \cite{13} and Heisenberg's paper of the July \cite{4} are:
 \\
 
 \noindent
{\bf I. Born-Jordan}: All elementary dynamics is discontinuous; there are no continuous orbits and virtual oscillators; radiation is quantized; the interaction of atoms and radiation occurs by discontinuous emission and absorption of photons.
\\
 {\bf Heisenberg's re-interpretation:} All elementary processes are continuous in space and time; radiation is classical; the emission of continuous radiation is due to virtual oscillators of the corresponding frequency.
\\

\noindent
 {\bf II. Born-Jordan} determine Einstein's probabilities for spontaneous emission and field induced emission and absorption of photons. The probabilities determine the intensities of spectral lines.
 \\ 
 {\bf Heisenberg's re-interpretation: } He relies on  Bohr-Sommerfeld quantization to determine quantized energies of virtual oscillators. This, in his opinion,  constitutes the ``integration of the equations of motion''. Virtual oscillator amplitudes determine intensities of spectral lines.
 \\
 
 \noindent
\\{\bf III. Born-Jordan} represent the atomic dipole moment by ``quantum vectors'' $\mathcal A(n,n-\tau)$, which are identical to the future ``matrix elements''. Discontinuous action intervals $\tau h$ correspond to emission of photons of energy $\epsilon_{photon} =\epsilon_n -\epsilon_{n-\tau}$. 
\\
{\bf Heisenberg's re-interpretation:} He adopts and re-interprets Born-Jordan's quantum vectors $\mathcal A(n,n-\tau)$; he restores continuity in time by a multiplicative phase factor $e^{2\pi i\;\nu (n, n-\tau)t}$, devised to fulfill Bohr's frequency condition $h\nu (n,n-\tau)= \epsilon_n - \epsilon_{n-\tau}$. The $\mathcal A(n,n-\tau) e^{2\pi i\;\nu (n, n-\tau)t}$ are interpreted as representations of position coordinates of virtual oscillators.
\\ 

\noindent {\bf IV. Born-Jordan} motivate the elimination of continuous variables by {\it ``The true laws of nature are determined only by such quantities, which are observable in principle''.\footnote{\it ``Ein Grundsatz von grosser Tragweite und Fruchtbarkeit besagt, dass in die wahren Naturgesetze nur solche Gr\"{o}\ss en eingehen, die prinziplell beobachtbar, feststellbar sind."}}
Nature is discontinuous at the elementary quantum scales; continuous variables loose their physical significance. 
\\
{\bf Heisenberg's re-interpretation:} He adopts the Born-Jordan principle of retaining only observable quantities, but re-interprets the reason for doing so. Although he still maintains that continuous electronic orbits and virtual oscillators constitute the underlying subatomic physics, he declares the subatomic dynamics to be  ``invisible in principle''. 
 This ``invisibility in principle'' of subatomic orbits and oscillators is presented as ad hoc postulate, without further justification.
 \\
 
 Heisenberg himself was not satisfied with this ad hoc postulate, and during the following years he searched for supporting arguments. The ``indeterminacy paper'' of March 1927 \cite{26} contains his ``explanation''.
\\

 \begin{subsection}
{The ``Measurement Problem''}
\end{subsection}

When Born and Jordan published the commutation relations in September 1925 \cite{1}, neither Heisenberg nor Bohr recognized their physical significance. 
Heisenberg remained fully attached to the classical concepts of Bohr's old Quantum Theory; continuity in space-time of all physical processes constituted its basic assumption, and exact values of all physical quantities at all times were taken for granted. But how could the commutation relations be reconciled with classical concepts? In particular, the apparent incompatibility of precise values of position and momentum called for an explanation.
\\

Heisenberg's ``indeterminacy paper'' of 1927 \cite{26} seemingly provided the answer. In order to ``explain'' the physical content of commutation relations, he invented the ``measurement problem'': The measurement of a physical quantity $q$ should necessarily cause unavoidable and uncontrollable disturbances of its canonically conjugated partner $p$. 
For example:  Both particle position and momentum have exact values at all times; these exact values may be determined separately. But while the position of a particle may be determined exactly (and thereby known), the act of position measurement necessarily disturbs its momentum, thereby precluding its  simultaneous determination (or knowledge). The accent here is on ``{\bf necessarily''} disturbs, and on {\bf ``simultaneous determination''}; only the {\bf simultaneous} determination of canonically conjugated quantities should be prohibited by unavoidable and uncontrollable disturbances produced by measurements. This reasoning was extended to all other physical quantities; Heisenberg remained convinced that 
\\
{\it ``All notions, which are used for the description of mechanical systems in classical theory, may be defined exactly also for atomic processes, in analogy to the classical notions''.}\footnote{\it ``Alle Begriffe, die in der klassischen Theorie zur Beschreibung eines mechanischen Systems verwendet werden, lassen sich auch f\"{u}r atomare Vorg\"{a}nge analog den klassischen Begriffen exakt definieren.''}
\\
While Born's discontinuous quantum physics contains {\bf quantum uncertainties} of all physical quantities as constitutive elements, Heisenberg's interpretation replaces quantum uncertainties by {\bf indeterminacies} resulting from experimental disturbances. 
\\

Heisenberg's ``explanation'' of the physical content of commutation relations is fundamentally wrong. Of course, many real measurements do disturb the system to be measured; that is true for many experiments in  classical physics and remains so for quantum physics. But Heisenberg's claim, that a measurement {\bf necessarily} disturbs the system to be measured, is false. 
Particle position measurements of atoms in crystalline materials provide the crucial example. Diffraction experiments are the standard method. Photons or neutrons or electrons are scattered off the crystal, the observed Bragg peaks provide the information necessary to determine the atomic positions. Bragg scattering processes are purely elastic; they not only leave the atomic positions unchanged, they also do not change their momenta.  The momentum transfer from the scattered particles (e.g. neutrons or photons) to the crystal is absorbed by the rigid crystal, while the momenta of individual  atoms bound in the crystal remain unchanged. Of course, inelastic scattering processes changing atomic momenta are possible, too; they are part of background contributions in addition to Bragg peaks.\footnote{Further details about Bragg scattering, background contributions, and measurements of position uncertainties are contained the appendix of ref.\cite{5}}
\\

Quantum uncertainties, not indeterminacies, are constitutive elements of quantum physics. Heisenberg's fundamental error invalidates all further conclusions, which he invokes to justify the retention of classical concepts.
\\

In order to quantify the indeterminacies, Heisenberg attributes dual properties, i. e. particle {\bf and} wave character, to individual photons and other particles; wavelength $\lambda$ and momentum $p$ are related by $\lambda = h/p$. In order to measure the position of a particle $X$, photons (or other particles) are scattered off particle $X$. The position indeterminacy of particle $X$ should be given by the wavelength $\lambda$ of the photons. Due to the Compton effect, the scattering process then should cause a momentum disturbance of particle $X$ of order $p=h/\lambda$. The resulting product of position and momentum indeterminacies then should be of order $h$. 
\\

Heisenberg accepts that Matrix Mechanics describes discontinuous quantum transition; but, in contrast to Born, he maintains the space-time continuum of Newtonian physics. He  attributes the origin of discontinuities to disturbances caused by measurements. Thereby, {\it ``neither the mathematical scheme of Quantum Mechanics requires a revision, nor is a revision of space-time geometry for small distances and times necessary''.}\footnote{\it ``das mathematische Schema der Quantenmechanik [wird] keiner Revision bed\"{u}rfen; ebensowenig wird eine Revision der Raum--Zeit Geometrie f\"{u}r kleine R\"{a}ume und Zeiten notwendig sein.''} 
\\

Furthermore, the retention of classical continuity in space and time required a justification for the statistical nature of quantum transitions. Again, the measurement problem is called to the rescue.  
{\it ``The fact that Quantum Theory may only provide the probability of electron positions (in the 1S-state for example) may, according to Born and Jordan, be viewed as characteristic and statistical elements of Quantum Theory in contrast to classical theory. But we might also state, as Dirac does, that statistics is introduced by our experiments''.}\footnote{\it ``Darin, dass in der Quantentheorie zu einem bestimmten Zustand, z. B.
1 S, nur die Wahrscheinlichkeitsfunktion des Elektronenortes angegeben
werden kann, mag man mit Born und Jordan einen charakteristisch
statistischen Zug der Quantentheorie im Gegensatz zur klassischen Theorie
erblicken. Man kann aber, wenn man will, mit Dirac auch sagen, dass
die Statistik durch unsere Experimente hereingebracht sei.''}
Heisenberg is in accord with Dirac, he specifies: {\it ``We did not assume that Quantum Theory is an essentially statistical theory in contrast to classical theory.....The sharp formulation of the causality law, 'if we know the present exactly, we are able to calculate the future', is not wrong due to the second part of the sentence, but because the precondition is wrong''.}\footnote{\it ``Dass die Quantentheorie im Gegensatz zur klassischen eine wesentlich
statistische Theorie sei in dem Sinne, dass aus exakt gegebenen Daten nur
statistische Schl\"{u}sse gezogen werden k\"{o}nnten, haben wir nicht angenommen.... An
der scharfen Formulierung des Kausalgesetzes: `Wenn wir die Gegenwart
genau kennen, k\"{o}nnen wir die Zukunft berechnen', ist nicht der
Nachsatz, sondern die Voraussetzung falsch''.} The statistical outcome thereby is attributed to imprecise knowledge of initial conditions. This argument applies equally to classical physics, which Heisenberg readily admits: {\it ``This would not be different in classical theory''.}\footnote{\it ``Dies w\"{a}re in der klassischen Theorie keineswegs anders.''}
\\

Imprecise knowledge of initial conditions is invoked to retain the concept of classical electron orbits. Starting from some initial conditions, where both $p$ and $q$ are supposed to be known with some indeterminacy, Heisenberg claims that 
\\{\it ``within the limits of the indeterminacies, the values of $q$ and $p$ obey classical equations of motion, as can be deduced directly from the quantum mechanical laws ${\bf \dot p = - \frac{\partial H}{\partial q}}$; ${\bf \dot q = \frac{\partial H}{\partial p}}$. As mentioned, the trajectory may only be computed statistically from the initial conditions, a fact which may be considered to result from the essential indeterminacy of the initial conditions''.}\footnote{\it ``die Werte von $p$ und $q$ innerhalb dieser Genauigkeitsgrenzen den klassischen  Bewegungsgleichungen Folge leisten, kann direkt aus den quantenmechanischen Gesetzen 
${\bf \dot p = - \frac{\partial H}{\partial q}}$; ${\bf \dot q = \frac{\partial H}{\partial p}}$
geschlossen werden.  Die Bahn kann aber, wie gesagt, nur statistisch aus den Anfangsbedingungen berechnet werden, was man als Folge der prinzipiellen Ungenauigkeit der Anfangsbedingungen betrachteten kann.''}
\\ 
Although these remarks suggest that, in Heisenberg's opinion, there should exist an underlying world, which is deterministic, he declares such speculations to be meaningless: 
\\
{\it ``Physics should merely provide a formal description for relations between observations''.}\footnote{\it ``Die Physik soll nur den Zusammenhang der Wahrnehmungen formal beschreiben.''}
\\

 This remark indicates the different philosophical attitudes of Born and Heisenberg concerning the question: ``What should a physical theory in general, and Quantum Theory in particular, achieve?'' Born's aim had been a logically consistent {\bf understanding} of quantum physics; Heisenberg is merely aiming at a formal description for relations between observations.  
 \\

Pursuing this objective, Heisenberg was forced into one further ad hoc hypothesis: The ``reduction of the wave packet''. Describing the initial position indeterminacy of an electron at some initial time by a wave packet, the wave packet should, according to Heisenberg's own reasoning, spread out in space with increasing time. Experimentally, however, single electrons are only observed as particles, not as extended waves; Heisenberg's ``explanation'' again involves the measurement problem: {\it ``Every determination of position reduces the wave packet to its original size''.}\footnote{\it ``Jede Ortsbestimmung reduziert das Wellenpaket wieder auf seine urspr\"{u}ngliche Gr\"{o}\ss e.''} This reduction of the wave packet is a special version of the general ``collapse of the wave function''. If, as is assumed by the Copenhagen interpretation, the time dependent Schr\"{o}dinger equation \cite{4} describes the continuous evolution of a particular physical system, then its mathematical form predicts the evolution from some particular initial state into a superposition of many different physical states at some later time, in contradiction to experimental observations. Here again the ``measurement problem'' is claimed to provide the explanation: The act of measurement causes the ``wave function collapse'' to the state actually observed. The collapse itself remains unexplained, which really means that this type of ``explanation'' does not explain anything. But that, according to Heisenberg's and Bohr's philosophy, does not constitute a problem, since {\it ``Physics should merely provide a formal description for relations between observations''.}\footnote{\it ``Die Physik soll nur den Zusammenhang der Wahrnehmungen formal beschreiben.''}
\\

\begin{subsection}
{The  Complementarity Principle}
\end{subsection}

In April 1928, Bohr published a review \cite{27}, essentially in agreement with Heisenberg. 
The space-time continuum of Newtonian physics and the conviction that all physical quantities have exact values at all times remained as prerequisites for all further conclusions. Slight differences between Heisenberg and Bohr are restricted to the emphasis attributed to discontinuities or wave properties. 
While Heisenberg considered the particle concept and discontinuities caused by measurements to be of primary importance; Bohr emphasized the wave character of particles. During the years preceding the discovery of the new quantum laws, Bohr had steadfastly rejected Einstein's photon concept of radiation; diffraction phenomena, in Bohr's view, provided definite proof of the wave character of radiation. Even if the discovery of the Compton effect forced Bohr to admit the existence of light quanta, he persisted; particle and wave character of photons should not be considered to be mutually exclusive, but ``complementary''. Observation of reflection maxima in scattering experiments of electrons off crystalline surfaces ``indisputably demonstrated'' the wave character of electrons and other particles as well.
\\

Particle-wave duality, elevated to ``Complementarity Principle'', played the crucial role in Bohr's interpretation of quantum physics. Bohr maintained that the classical concepts of the old Quantum Theory should still provide the framework for the description of all observations. He argued that all measurements have to rely on macroscopic, i.e. necessarily classical, instruments. Measuring the properties of a quantum system, e.g. an atom, measuring apparatus and atom should be viewed as one interrelated total system; no separate reality should be attributed to the atom alone. And since the measuring instrument has to be described classically, Bohr concluded, that the description of atomic properties, too, must be in terms of classical concepts. The limits of obtainable accuracy in measuring physical quantities of quantum systems should be determined by the limits of accuracy  of the combined system, consisting of quantum object and measuring apparatus. Applying this reasoning to the measurement of electron position by optical means, Bohr invoked the resolution limit of classical microscopes as limit to the accuracy of electron position. Assuming that arbitrarily accuracy may be achieved using correspondingly shorter wave lengths, the particle character of photons should cause correspondingly large momentum transfers due to the Compton effect, leading to larger and larger inaccuracies of momentum. 
\\

Bohr relies on two hypotheses, which are both fundamentally wrong. 
\\
I) Classical instruments may be made arbitrarily precise; they may determine exact values of physical quantities (e.g. position), but do not allow the simultaneous determination of its canonically conjugated partner (here momentum).
\\
This was Heisenberg's main argument, when he invented the measurement problem; Heisenberg's fundamental error has already been pointed out above. 
\\
Bohr's second argument:
\\
II) The limit of accuracy is given by the instrumental resolution.
\\
This second argument is not only wrong, but in complete contradiction to experimental practice. Although progress in experimental methods was often essential in understanding quantum phenomena, limited instrumental resolution has not prevented conclusions about scales, far beyond the accuracy of the instruments themselves. Progress in quantum physics has rather been obtained from {\bf new insight}, which provided a {\bf consistent understanding} of observations. Thus, information about subatomic length scales is obtained from high momentum transfer scattering experiments, relying on detectors whose accuracy is many orders of magnitude worse. 
\\

 But a consistent understanding of quantum physics was not Bohr's objective, nor Heisenberg's \cite{26}: {\it ``Physics should merely provide a formal description for relations between observations.''}  And if mutually exclusive notions had to be invoked to describe different observations, {\it `` it is a question of convenience at what point the concept of observation involving the quantum postulate with its inherent 'irrationality ' is brought in......The two views of the nature of light are rather to be considered as different attempts at interpretation of experimental evidence''}.\cite{27}
\\

\begin{subsection}
{The Consolidation of the Copenhagen Interpretation}
\end{subsection}

By 1930, Wave Mechanics superseded Matrix Mechanics almost completely; Quantum Theory was identified with 'Schr\"{o}dinger equation'. The space-time continuum of Newtonian physics as prerequisite for the understanding of quantum physics was taken for granted. The Copenhagen interpretation had gained widespread recognition; classical concepts, in particular continuity in space and time, remained to be its central doctrine. The book by Born and Jordan ``Elementare Quantenmechanik" \cite{18}, published in 1930, constituted a belated attempt to stem the tide. Right away on the first page of \textsection 1  we find:
\\

{\it ``There can be no question of an ``explanation" of the unfamiliar quantum laws by means of reduction to classical concepts; on the contrary the fundamental and primary character of the basic quantum theoretical assumptions emerged clearly only due to new developments. Progress consists precisely in abandonment of the remains of classical views; as a result a self-contained theory emerged, which allows to describe all atomic processes consistently and which contains the classical theory as special limit".}\footnote{\it ``Es ist also keine Rede von einer ``Erkl\"{a}rung" der fremdartigen Quantengesetze durch eine Zur\"{u}ckf\"{u}hrung auf klassische Vorstellungen; im Gegenteil ist der prinzipielle, prim\"{a}re Charakter der quantentheoretischen Grundannahmen erst durch die neuere Entwicklung klar zum Vorschein gekommen. Der Fortschritt besteht gerade im Abstreifen der Reste klassischer Betrachtungsweise; dadurch ist eine in sich geschlossene Theorie entstanden, die alle Atomvorg\"{a}nge widerspruchsfrei zu beschreiben gestattet und die klassische Theorie als speziellen Grenzfall enth\"{a}lt."}
\\

Pauli, who had become the most fervent advocate of the Copenhagen interpretation, reacted by a scathing review. \cite{28} Start and finish are outright ridicule:
\\
 {\it "The book is the second volume of a series, which explains aim and meaning of the n'th volume by the virtual existence of the (n+1)'th volume........ The features of the book as far as print and paper are concerned are excellent".}
\\
 In between Pauli criticizes the algebraic methods as
\\
 {\it "inhibiting the insight into scope and internal logic (!) of the theory....such as the statistical interpretation of Quantum Mechanics (!)".}
\\
 Condescending advice is given to Born and Jordan (authors of the fundamental laws of Quantum Theory!) not to delve into statements of principle, such as the postulate 
\\
{\it ``to represent each physical quantity by a matrix"},
\\
 and leave the interpretation to the true owners of the understanding, i.e. the followers of the Copenhagen interpretation. 
\\
{\it ``The meaning of such a ``representation" of a physical quantity in reality can be understood only due to later conclusions"},
\\
 i.e. Heisenberg's explanation of ``indeterminacies'', Bohr's stationary states, collapse of the wave function, complementarity, and particle-wave duality. Therefore Born and Jordan should
\\
 {\it ``restrict the theory to the methods of measurements of particle position and momentum or of energy eigenvalues of stationary states and to the postulates of possible measurements obtained from the general wave-particle experience."}
\\

If Quantum Theory is widely not understood until today, it is due to the ``success'' of the Copenhagen interpretation. Majority opinion attributed the Nobel prize for 1932 to Heisenberg ``For the creation of quantum mechanics''. During the following decades, the Copenhagen interpretation constituted the basis for almost all textbooks on Quantum Mechanics. This heritage dominates teaching of elementary Quantum Theory until today.
\\

\begin{section}
{Conclusion}
\end{section}

The scientific revolutions of the first quarter of the 20'th century, Relativity Theory and Quantum Theory,  both rest on fundamental principles, imposed by universal constants. Relativity Theory is based in on the velocity of light $c$ being a universal constant. Velocity of light $c$ not being infinite requires a redefinition of space-time on large and cosmological scales.  Einstein recognized that there is no space-time given ``a-priori'', independent of all empirical facts. Physical notions of space and time are related to measurements, and, light velocity not being infinite, implied that measurements of spacial distances and time intervals depend on the reference system from which measurements are performed.  Einstein required that this should be reflected by the basic laws of Relativity Theory.\\

The primary aim of this paper is to show that quantization of action in terms of a finite, i.e. non vanishing, universal quantity $h$ requires a redefinition of space-time on the elementary quantum scale. While Relativity Theory still retained the continuum, albeit reference dependent, Born recognized that the continuum as prerequisite to all understanding must be replaced by a discrete manifold on the elementary quantum scale. Action variables may only change by integer multiples of Planck's quantum of action, requiring all other physical quantities to change by finite steps as well. Furthermore, discreteness of all events in nature eliminates the justification for determinism, implied by the differential equations of classical physics. All elementary processes in nature are discrete and governed by statistical laws. And, similar to Einstein, Born required that this should be reflected by the basic laws of Quantum Theory. Consistent with the discrete character of nature, Born and Jordan derived the basic laws of Quantum Theory in discrete mathematical forms, Matrix Mechanics. 
\\

Born-Wiener \cite{19} and Schr\"{o}dinger \cite{3,4} soon replaced matrices by field theoretical forms, easier to handle mathematically. But, in addition to easier mathematics, field theoretical forms led to fundamental misunderstandings.  The variables ${\bm r}$ and $t$ of Schr\"{o}dinger's wave function suggested Newtonian space-time coordinates; the continuous variable $t$ was associated with continuous variation of physical quantities in time. This misunderstanding was enforced, when Matrix and Wave Mechanics were shown to be equivalent. Born on one side and Schr\"{o}dinger on the other had opposing views of this equivalence, which may be distinguished by {\bf ``Born equivalence''} vs. {\bf ``Schr\"{o}dinger equivalence''}. 
The latter became official doctrine, finalized by the supposed equivalence of `Schr\"{o}dinger representation' (states are time dependent) and `Heisenberg representation' (operators are time dependent, eq. 10). In section(5) (Time in Quantum Physics) the difference between time $t$ as external parameter and time for closed systems has been highlighted. 
In \textsection 22 of the book Elementare Quantemnechanik (ref. \cite{18}), Born specifies that time $t$ in the Heisenberg representation is external parameter, not directly relevant for the temporal behavior of individual quantum systems. But Born's insistence was completely lost by the scientific community. The letter $t$ was interpreted in Schr\"{o}dinger's sense as Newtonian time. 
\\

Discontinuous changes of action variables by integer multiples of Planck's constant $h$ represent the key to understanding Quantum Theory. The mathematical implementation of this basic requirement resulted in commutation relations for canonically conjugated physical quantities. All further conclusions  are direct consequences of this quantization condition. Most importantly: There is no continuous physical time.
Concerning  origin and physical significance of commutation relations, the scientific community widely ignored Born's reasoning. Field theoretical representations and the advent of quantum field theory consolidated the general conviction, that Quantum Theory retains Newtonian space-time notions, physical processes occurring continuously {\bf in} space-time. But changing mathematical representations from discrete matrix calculus to operators and  functions of continuous variables does not alter the physical content. All of physics is contained in matrix elements, and matrix elements are independent of the particular representation used to obtain them. 
Actually, a close look at the application of Quantum Theory to the analysis of experimental results reveals that ``Born equivalence'' does constitute general practice! Typically, the original assumption of continuity in time is abandoned at the very end: Experimental evidence is incompatible with what a continuum theory would predict. To obtain accord with observed facts, the application of Born's statistical interpretation becomes necessary, which implicitly means that discreteness has been reintroduced through the back door. Any type of ``collapse of the wave function'' is equivalent to Born's statistical interpretation and the recognition of discreteness. While typical language upholds ``Schr\"{o}dinger equivalence'', final practice amounts to ``Born equivalence''. 
\\

A logically consistent understanding of Quantum Theory is obtained by going back to the origin: Max Born not only provided the first representation of its fundamental equations (together with Pascual Jordan \cite{1}), he also recognized the basic principles of quantum physics. Nature is discrete and statistical at the elementary level: {\bf ``Action variables may only change by integer multiples of $h$, requiring all other physical quantities to change by finite steps as well''}.
\\
\vspace{10mm}

\noindent
{\bf Acknowledgment:}  Many thanks to Efim I. Kats, Maarten Wegewijs, and Nikolai Sinitsyn for helpful comments.
\vspace{20mm}


\end{document}